\documentclass[aip, apl, reprint]{revtex4-1}

\pdfoutput=1 
\usepackage{graphicx}
\usepackage{color,amsmath}
\usepackage[normalem]{ulem}

\usepackage[colorlinks,linkcolor=blue,citecolor=blue,urlcolor=blue]{hyperref}
\usepackage{siunitx}
\usepackage{comment}
\usepackage{layouts}

\usepackage{amssymb}
\usepackage{hyperref}
\usepackage[english]{babel}
\usepackage[utf8]{inputenc}
\begin{document}
\title{Gate-Tunable Quantum Dot in a High Quality Single Layer MoS$_{\mathrm{2}}$ Van der Waals Heterostructure}

\author{Riccardo Pisoni}
\email{pisonir@phys.ethz.ch}
\affiliation{Solid State Physics Laboratory, Department of Physics, ETH Z{\"u}rich, 8093 Z{\"u}rich, Switzerland}

\author{Zijin Lei}
\affiliation{Solid State Physics Laboratory, Department of Physics, ETH Z{\"u}rich, 8093 Z{\"u}rich, Switzerland}

\author{Patrick Back}
\affiliation{Institute of Quantum Electronics, Department of Physics, ETH Z{\"u}rich, 8093 Z{\"u}rich, Switzerland}

\author{Marius Eich}
\affiliation{Solid State Physics Laboratory, Department of Physics, ETH Z{\"u}rich, 8093 Z{\"u}rich, Switzerland}

\author{Hiske Overweg}
\affiliation{Solid State Physics Laboratory, Department of Physics, ETH Z{\"u}rich, 8093 Z{\"u}rich, Switzerland}

\author{Yongjin Lee}
\affiliation{Solid State Physics Laboratory, Department of Physics, ETH Z{\"u}rich, 8093 Z{\"u}rich, Switzerland}

\author{Kenji Watanabe}
\affiliation{National Institute for Material Science, 1-1 Namiki, Tsukuba 305-0044, Japan}

\author{Takashi Taniguchi}
\affiliation{National Institute for Material Science, 1-1 Namiki, Tsukuba 305-0044, Japan}


\author{Thomas Ihn}
\affiliation{Solid State Physics Laboratory, Department of Physics, ETH Z{\"u}rich, 8093 Z{\"u}rich, Switzerland}

\author{Klaus Ensslin}
\affiliation{Solid State Physics Laboratory, Department of Physics, ETH Z{\"u}rich, 8093 Z{\"u}rich, Switzerland}

\date{\today}

\begin{abstract}
We have fabricated an encapsulated monolayer MoS$_{\mathrm{2}}$ device with metallic ohmic contacts through a pre-patterned hBN layer. In the bulk, we observe an electron mobility as high as 3000 cm$^{\mathrm{2}}$/Vs at a density of 7 $\times$ 10$^{\mathrm{12}}$ cm$^{\mathrm{-2}}$ at a temperature of 1.7 K. Shubnikov-de Haas oscillations start at magnetic fields as low as 3.3 T. By realizing a single quantum dot gate structure on top of the hBN we are able to confine electrons in MoS$_{\mathrm{2}}$ and observe the Coulomb blockade effect. By tuning the middle gate voltage we reach a double dot regime where we observe the standard honeycomb pattern in the charge stability diagram.
\end{abstract}

\maketitle

Contrary to graphene, in monolayer molybdenum disulfide (MoS$_{\mathrm{2}}$) inversion symmetry is broken. This, together with the presence of time-reversal symmetry, endows single layer MoS$_{\mathrm{2}}$ with individually addressable valleys in momentum space at the K and K${\mathrm{'}}$ points in the first Brillouin zone.~\cite{xiao_coupled_2012,chhowalla_chemistry_2013,xu_spin_2014,kormanyos_monolayer_2013} This valley addressability facilitates the momentum state of electrons to be used for novel qubit architectures. Recent theoretical works have been exploring the possibility of using spin and valley states of gate-defined quantum dots in 2D MoS$_{\mathrm{2}}$ as quantum bits.~\cite{kormanyos_spin-orbit_2014,novoselov_2d_2016,loss_quantum_1998}
In this manuscript, we describe the observation of Coulomb blockade in single and coupled dot in a high quality single layer MoS$_{\mathrm{2}}$. The high electronic quality of our monolayer MoS$_{\mathrm{2}}$ results in the observation of Shubnikov-de Haas oscillations (SdHO) occurring at magnetic fields as low as 3.3 T. The 2DEG in the MoS$_{\mathrm{2}}$ can be electrostatically depleted below the gate pattern with resistance values exceeding the resistance quantum h/e$^{\mathrm{2}}$ by orders of magnitude. We observe Coulomb blockade resonances close to pinch-off indicating single electron tunneling in and out of the dot. By adjusting the gate voltages, we are able to tune the electrostatic landscape inside the dot and to form a double dot system within a single dot gate structure.~\cite{zhang_electrotunable_2017,song_gate_2015,wang_engineering_2016,epping_quantum_2016}
\vspace{6pt}
\begin{figure*}[]
	\includegraphics[width=\textwidth]{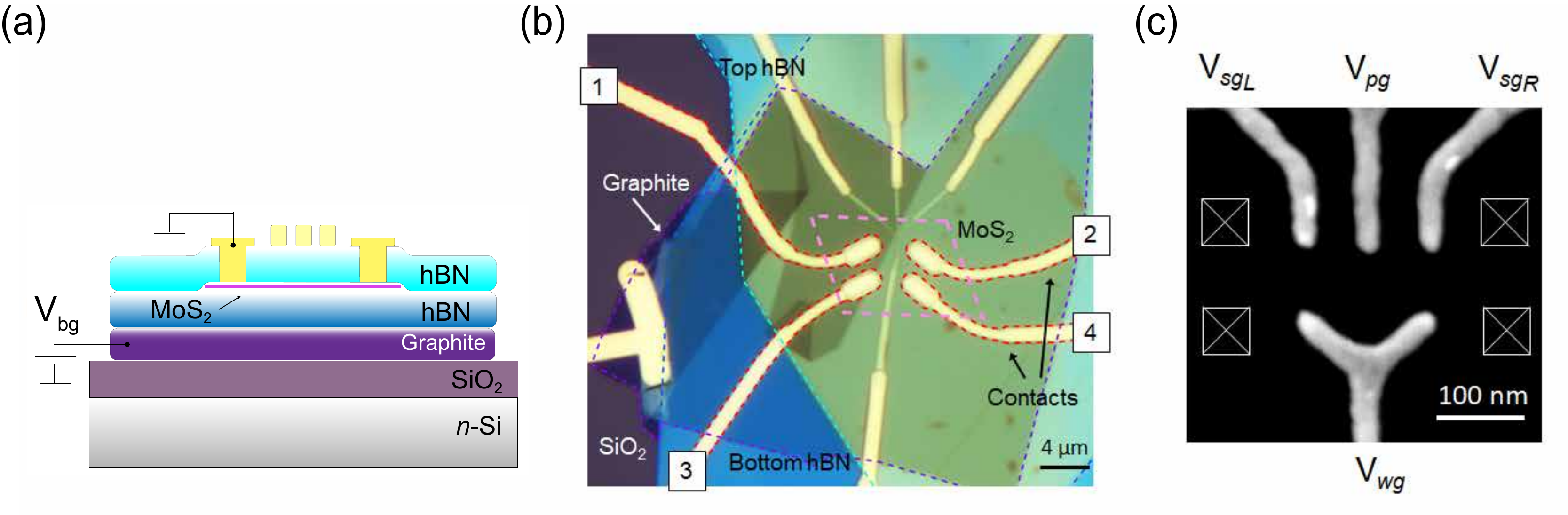}
	\caption{\textbf{(a)} Cross-sectional schematic of the MoS$_{\mathrm{2}}$-based van der Waals heterostructure. The top hBN has been patterned by means of e-beam lithography and reactive ion etching, before being picked-up. Ti/Au metallic leads have been evaporated directly in contact with the MoS$_{\mathrm{2}}$ layer. \textbf{(b)} Optical micrograph of the device. A monolayer MoS$_{\mathrm{2}}$ (pink dashed lines) has been encapsulated between two hBN flakes. We use graphite as a backgate and Ti/Au contacts (numbered 1-4) evaporated on top of the MoS$_{\mathrm{2}}$ layer. \textbf{(c)} SEM image of the single dot device. The structure is realized on top of hBN, bright regions correspond to the metallic electrodes. For the Coulomb blockade measurements (Fig.~\ref{fig3}), V$_{\mathrm{wg}}$ = -8 V.} 
	\label{fig1}
\end{figure*} 

In Fig.~\ref{fig1}(a), we show the schematic of a monolayer MoS$_{\mathrm{2}}$ ($\sim$ 0.7 nm thick) encapsulated between two hexagonal boron nitride (hBN) layers. The measured MoS$_{2}$ flake was exfoliated from natural bulk crystal (SPI supplies). The bottom hBN layer is $\sim$ 30 nm thick and works both as an atomically flat substrate and as a dielectric that isolates the MoS$_{\mathrm{2}}$ from a graphite backgate.  The graphite gate enables us to control electrostatically the electron density in MoS$_{\mathrm{2}}$ with the voltage V$_{\mathrm{bg}}$. The layers thicknesses were determined by atomic force microscopy (AFM). The top hBN layer ($\sim$50nm thick) has been pre-patterned using E-beam lithography and reactive ion etching.~\cite{wang_electronic_2015} This enables us to evaporate metallic contacts (Ti/Au) on top of the MoS$_{\mathrm{2}}$  layer where the hBN has been etched away, without exposing the channel region to organic residues remaining from the fabrication process.~\cite{allain_electrical_2015} Prior to metal evaporation, the heterostructure has been annealed in forming gas (Ar/H$_{\mathrm{2}}$) at 300$^{\circ}\mathrm{C}$  for 30 minutes in order to remove most of the organic residues on top of the MoS$_{\mathrm{2}}$ contact regions and to achieve ohmic contact behavior. 5 nm of Ti and 65 nm of Au were deposited by means of electron-beam evaporation at a pressure of $\sim$ 2$\times$10$^{-8}$ mbar. This fabrication process enables us to realize ohmic contacts on monolayer MoS$_{\mathrm{2}}$ without the use of graphene or Co/h-BN electrodes as in previous works.~\cite{cui_low-temperature_2017,cui_multi-terminal_2015} Fig.~\ref{fig1}(b) shows the optical micrograph of the device where the monolayer MoS$_{\mathrm{2}}$ is outlined in pink. The MoS$_{\mathrm{2}}$ crystal was exfoliated and the heterostructure assembled in an argon environment (H$_{\mathrm{2}}$O and O$_{\mathrm{2}}$ levels $<$ 0.1 ppm).~\cite{bretheau_tunnelling_2017,pisoni_gate-defined_2017} The top hBN serves as a dielectric layer for the top gate structure. Fig.~\ref{fig1}(c) displays the scanning electron microscope (SEM) image of the top gate structure in which the voltages applied to the gates are labeled V$_{\mathrm{sgL}}$, V$_{\mathrm{pg}}$, V$_{\mathrm{sgR}}$ and V$_{\mathrm{wg}}$. To avoid electrostatic inhomogeneities, the gate structure has been deposited on a bubble-free region on top of the MoS$_{\mathrm{2}}$. The gate-defined lithographic dot radius is $\sim$ 70 nm.

\begin{figure*}[]
	\includegraphics[width=\textwidth]{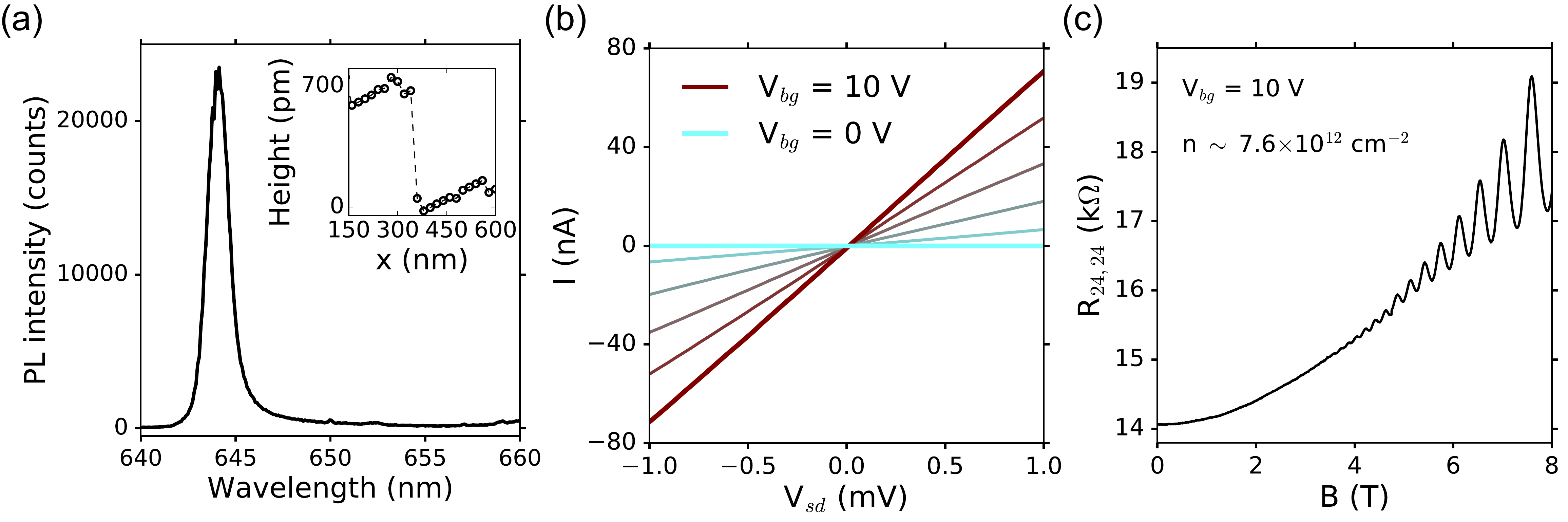}
	\caption{\textbf{(a)} PL spectrum for monolayer MoS$_{\mathrm{2}}$ at V$_{\mathrm{bg}}$ = -1 V. The PL peak shows a FWHM $\sim$ 4 meV and is centered at $\sim$ 1.925 eV as determined with a Lorentzian fit. We use a HeNe excitation laser. \textbf{(b)} Linear I V$_{\mathrm{sd}}$ curves are observed for V$_{\mathrm{bg}}$$\geq$1.5 V at T $\sim$ 1.7 K. Different colors correspond to different V$_{\mathrm{bg}}$. \textbf{(c)} Two-terminal resistance as a function of B field for a single layer MoS$_{\mathrm{2}}$ device measured at T $\sim$ 1.7 K and n $\sim$ 7.6 $\times$ 10$^{\mathrm{12}}$ cm$^{\mathrm{-2}}$. }  
	\label{fig2}
\end{figure*}

Fig.~\ref{fig2}(a) displays the measured photoluminescence (PL) spectrum at 4 K and the AFM height signal (inset) of the MoS$_{\mathrm{2}}$ flake. The shown PL spectrum was measured at V$_{\mathrm{bg}}$ = -1 V, where the sample is devoid of free electrons. We note that the PL signal is slightly distorted on the shorter wavelength side because of the longpass filter used to suppress the excitation HeNe laser. The Full Width at Half Maximum (FWHM) of the exciton resonance is $\sim$ 4 meV and the peak is centered at a wavelength of 644 nm corresponding to $\sim$ 1.925 eV. For a few layer sample the PL spectrum consists of multiple emission peaks and lower PL intensity, we then conclude that our sample is a single layer MoS$_{\mathrm{2}}$.~\cite{mak_atomically_2010} The AFM height signal shows a step height of $\sim$ 0.7 nm, in agreement with previous single layer MoS$_{\mathrm{2}}$ characterizations.~\cite{radisavljevic_single-layer_2011} Fig.~\ref{fig2}(b) shows the low temperature (T $\sim$ 1.7 K) current (I) flowing through the device as a function of the voltage (V$_{\mathrm{sd}}$) applied between two electrodes (2 and 4 in Fig.~\ref{fig1}(b)) at various V$_{\mathrm{bg}}$ and zero voltage between patterned top gates and 2DEG. Linear I-V$_{\mathrm{sd}}$ curves are observed for V$_{\mathrm{bg}}$$\geq$1.5 V indicating ohmic contact behavior. The resistance decreases with increasing V$_{\mathrm{bg}}$ as expected for an n-type semiconductor.~\cite{cui_multi-terminal_2015} In order to estimate the quality of our device we performed magnetotransport measurements at T $\sim$ 1.7 K, using standard lock-in techniques at 80.31 Hz. Fig.~\ref{fig2}(c) shows the resistance R$_{\mathrm{24,24}}$ as a function of magnetic field B, at the Hall density n $\sim$ 7.6 $\times$ 10$^{\mathrm{12}}$ cm$^{\mathrm{-2}}$. We observe the appearance of SdHO at B $\sim$ 3.3 T, yielding a lower limit for the mobility of 3000 cm$^{\mathrm{2}}$/Vs.

\begin{figure*}[]
	\includegraphics[width=\textwidth]{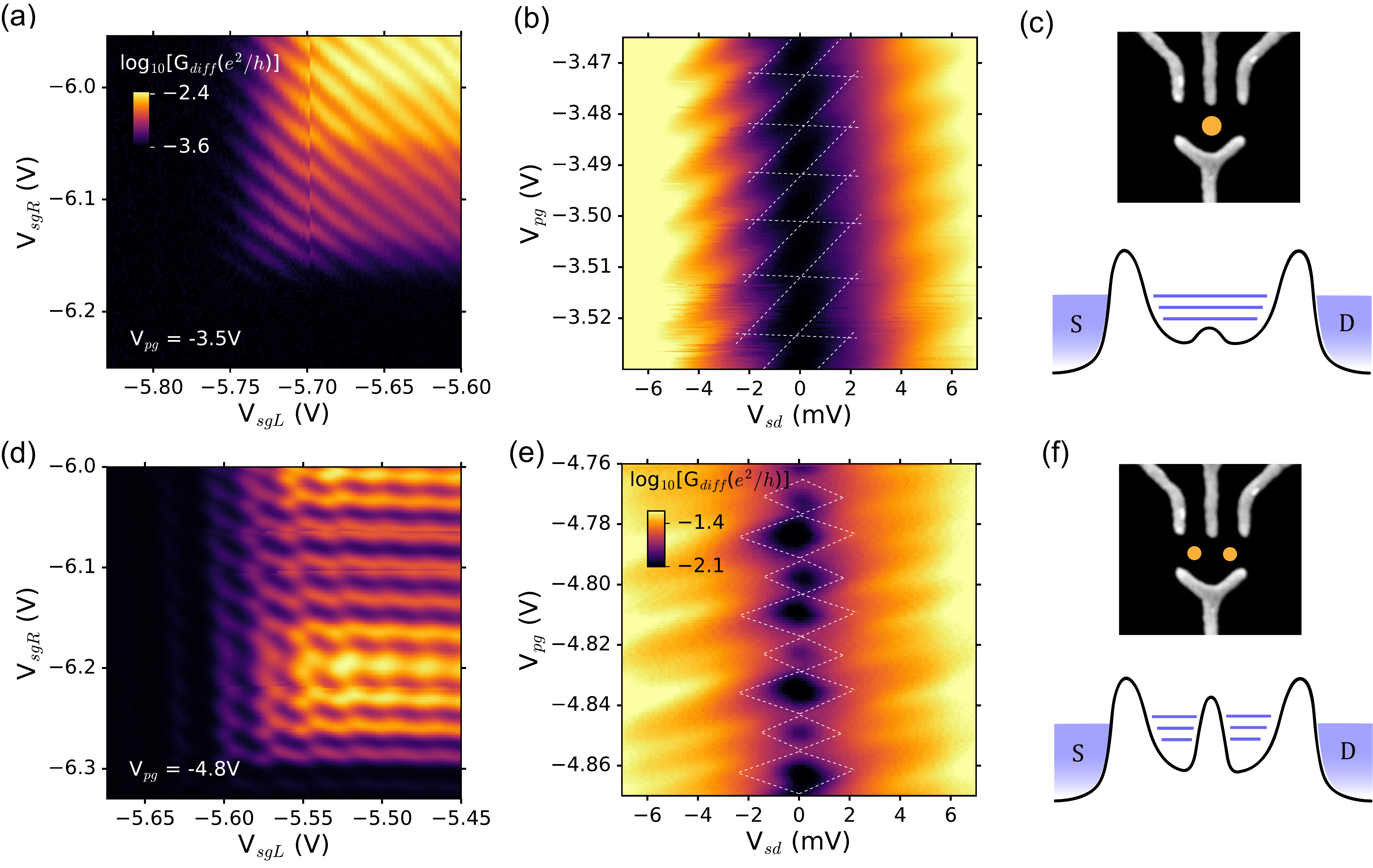}
	\caption{\textbf{(a)(d)} Logarithm of the single and double dots differential conductance as a function of the barrier gate voltages V$_{\mathrm{sgR}}$ and V$_{\mathrm{sgL}}$ measured at T $\sim$ 1.7 K and V$_{\mathrm{bg}}$ = 3.4 V. Dark regions indicate low conductance, bright represent high conductance. The color scale is the same for both figures. \textbf{(b)(e)} Coulomb blockade diamonds in the single and double dots regime. Differential conductance in logarithmic scale is displayed as a function of V$_{\mathrm{pg}}$ at T $\sim$ 1.7 K and V$_{\mathrm{bg}}$ = 3.4 V. The color scale is the same for both figures. In \textbf{(b)} V$_{\mathrm{sgL}}$ = -5.7 V, V$_{\mathrm{sgR}}$ = -6.1 V. In \textbf{(e)} V$_{\mathrm{sgL}}$ = -5.6 V, V$_{\mathrm{sgR}}$ = -6.1 V. \textbf{(c)(f)} Electrostatic potential landscape without and with the middle barrier potential that allows us to monotonically tune our device from single to double dot behavior.}  
	\label{fig3}
\end{figure*}

The 2DEG in the MoS$_{\mathrm{2}}$ can be locally depleted by applying negative voltages to the top gate structure. This structure allows us to confine electrons in a region of space small enough to observe the Coulomb blockade effect.~\cite{kouwenhoven_electron_1997} Fig.~\ref{fig3} displays the charge stability diagram and the transport spectroscopy of the quantum dot, at T $\sim$ 1.7 K and V$_{\mathrm{bg}}$ = 3.4 V. By varying V$_{\mathrm{pg}}$ we can modify the electrostatic potential landscape inside the dot and tune the coupling strength in order to observe both single and double dot regimes. Fig.~\ref{fig3}(a, d) show the differential conductance on a logarithmic scale as a function of the left (V$_{\mathrm{sgL}}$) and right (V$_{\mathrm{sgR}}$) barrier gate voltages at V$_{\mathrm{pg}}$ = -3.5 V and -4.8 V respectively. When V$_{\mathrm{pg}}$ = -3.5 V, close to pinch off, we observe parallel resonances with intermittent Coulomb blockade for tunneling of electrons on and off the single dot (Fig.~\ref{fig3}(c)).  The lines are rather straight indicating that additional localized states possibly caused by disorder in MoS$_{\mathrm{2}}$ are of minor importance. The high quality of our monolayer MoS$_{\mathrm{2}}$ device yields a mean free path of $\sim$ 150 nm, slightly larger than the dot size. The measured conductance peaks exhibit a FWHM $\sim$ 5K$_{\mathrm{B}}$T as expected for thermally broadened peaks in the multilevel transport regime.~\cite{beenakker_theory_1991} From the peaks spacing along the V$_{\mathrm{sgL}}$ and V$_{\mathrm{sgR}}$ axis in Fig.~\ref{fig3}(a), we estimate the capacitance between the dot and the left and right barrier gates to be $\sim$ 6.7 aF and $\sim$ 5.9 aF respectively.~\cite{kouwenhoven_electron_1997} By tuning the gate voltage V$_{\mathrm{pg}}$ to more negative values, as shown in Fig.~\ref{fig3}(d), we observe a hexagonal array of lines which is the signature for the formation of two coupled quantum dots (Fig.~\ref{fig3}(f)).~\cite{van_der_wiel_electron_2002} According to the dimensions of the honeycomb pattern in Fig.~\ref{fig3}(d), the capacitances between the dot and the left and right barrier gates can be determined to be $\sim$ 5.8 aF and $\sim$ 4.8 aF respectively and the cross-capacitances to be $\leq$ 0.5 aF. Fig.~\ref{fig3}(b, e) display the differential conductance on a logarithmic scale as a function of the source-drain bias V$_{\mathrm{sd}}$ and V$_{\mathrm{pg}}$. In the single dot regime (Fig.~\ref{fig3}(b), we observe regular even-spaced Coulomb diamonds. From the height of the Coulomb diamonds along the V$_{\mathrm{sd}}$ axis, we estimate the charging energy to be $\sim$ 2 meV. By employing $Ec = \frac{e^{2}}{8\epsilon_{\mathrm{0}}\epsilon_{\mathrm{r}}r}$,~\cite{kouwenhoven_electron_1997} with $\epsilon_{\mathrm{r}}$ = 4 the relative permittivity of hBN,~\cite{young_electronic_2012} we estimate the radius of the dot to be $\sim$ 280 nm. The discrepancy between the estimated dot radius and the gate structure design can be assigned to the presence of the metallic gates and the adjacent source and drain leads that significantly affect the capacitance.~\cite{kouwenhoven_electron_1997} From the Coulomb blockade diamond measurements, we read the plunger gate lever arm $\alpha_{\mathrm{pg}}$ $\sim$ 0.19 as the ratio between the charging energy and the peaks spacing along the Vpg axis. In the double dot regime (Fig.~\ref{fig3}(e)), we observe the interchange of small and big Coulomb diamonds that, together with the honeycomb pattern in Fig.~\ref{fig3}(d), may indicate the presence of two quantum dots.~\cite{kouwenhoven_electron_1997}
\vspace{6pt}

In conclusion, we have developed a fabrication technique, that allows us to realize ohmic contacts on monolayer MoS$_{\mathrm{2}}$ by employing metallic gates evaporated directly on top of the MoS$_{\mathrm{2}}$ surface instead of using graphene electrodes or Co/h-BN contacts,~\cite{cui_low-temperature_2017,cui_multi-terminal_2015} without affecting the electronic quality of the 2DEG. We observe the appearance of SdHO at lower magnetic field compared to previous work,~\cite{cui_multi-terminal_2015} indicating an electron mobility exceeding 3000 cm$^{\mathrm{2}}$/Vs. By electrostatically depleting the MoS$_{\mathrm{2}}$ 2DEG we are able to confine electrons and observe Coulomb blockade effects in single and double dots. The samples are clean enough that the disorder-localized states do not affect the experimental observations. Gate-defined quantum dots formed in 2D semiconducting Transition Metal Dichalcogenides are the first step towards the realization of spin-valley qubits, filters and other intriguing valleytronic devices.~\cite{kormanyos_spin-orbit_2014,loss_quantum_1998,brooks_spin-degenerate_2017}

%
   
\begin{acknowledgments}
The authors acknowledge financial support from the Graphene Flagship, the EU Spin-Nano RTN network, and the National Center of Competence in Research on Quantum Science and Technology (NCCR QSIT) funded by the Swiss National Science Foundation. Growth of hexagonal boron nitride crystals was supported by the Elemental Strategy Initiative conducted by the MEXT, Japan and JSPS KAKENHI Grant Numbers JP15K21722.
\end{acknowledgments}

\bibliography{ref}
\end{document}